\documentclass[conference]{IEEEtran}

\IEEEoverridecommandlockouts

\ifCLASSOPTIONcompsoc
  \usepackage[caption=false,font=normalsize,labelfont=sf,textfont=sf]{subfig}
\else
  \usepackage[caption=false,font=footnotesize]{subfig}
\fi

\usepackage{mathtools,amssymb}

\usepackage{amsmath, nccmath}

\usepackage{cite}
\usepackage{bm}

\usepackage{algorithm}

\usepackage[noend]{algpseudocode}

\newcounter{MYtempeqncnt}

\newcommand{\Lagr}{\mathcal{L}}

\begin{document}

\title{Interference Regime Enforcing Rate Maximization for Non-Orthogonal Multiple Access (NOMA)\thanks{DISTRIBUTION A. Approved for public release: distribution unlimited.}}

\author{\IEEEauthorblockN{Tugba Erpek}
\IEEEauthorblockA{Intelligent Automation, Inc.\\
Rockville, MD, USA\\
terpek@i-a-i.com}
\and
\IEEEauthorblockN{Sennur Ulukus}
\IEEEauthorblockA{University of Maryland\\
College Park, MD, USA\\
ulukus@umd.edu}
\and
\IEEEauthorblockN{Yalin E. Sagduyu}
\IEEEauthorblockA{Intelligent Automation, Inc.\\
Rockville, MD, USA\\
ysagduyu@i-a-i.com}
}

\maketitle

\begin{abstract}
An interference regime enforcing rate maximization scheme is proposed to maximize the achievable ergodic sum-rate of the parallel Gaussian interference channels by enforcing very strong interference at receivers through power allocation whenever strong interference is observed. Applying successive interference cancellation (SIC) at the receivers, very strong interference can be completely eliminated in this case that enables non-orthogonal multiple access (NOMA). An optimization problem is formulated for ergodic rate maximization, where the conditions for creating very strong interference are included as additional constraints. Then, a generalized iterative waterfilling algorithm is derived to solve this optimization problem. The resulting power allocation scheme is compared in terms of the sum-rate with two other schemes when the interference is treated as noise with optimal or sub-optimal power allocation. The results show that the interference regime enforcing scheme provides major improvement to the sum-rate of parallel Gaussian interference channels and enables NOMA.
\end{abstract}

\begin{IEEEkeywords}
Parallel Gaussian interference channel, NOMA, waterfilling, interference cancellation, power allocation.
\end{IEEEkeywords}

\section{Introduction}
In an interference channel (IC), multiple transmitters communicate with their intended receivers on the same channel causing interference to each other. The IC capacity region has been investigated for \emph{weak}, \emph{strong} and \emph{very strong} interference regimes. Channel gains, noise power and transmit power allocation jointly determine the type of interference regime. When the  interference is weak, treating interference as noise achieves the sum-capacity, as shown in \cite{veeravalli}.  The sum-capacity of specific sub-classes of ergodic fading ICs was provided in \cite{SankarPoor}. The IC rate in  the  strong  interference  regime  was determined in  \cite{HanKobayashi}, where each user decodes the information transmitted from the other user. This scheme is computationally very expensive. As a result, single-user detection approaches have been mostly adopted at the receivers by treating the interference as noise \cite{ZhangCui,ErpekVTC}. 

In \cite{Carleial}, it was shown that interference does not necessarily reduce the channel capacity of a two-user IC when it is very strong, since very strong interference can be perfectly eliminated. Recently, non-orthogonal multiple access (NOMA) techniques have been developed for 5G systems, where multiple users are served simultaneously in the same resource block. NOMA schemes have also been applied to MIMO systems \cite{MIMONOMAPoor} and it was shown that the sum capacity is better for MIMO-NOMA compared to MIMO-orthogonal multiple access (MIMO-OMA) \cite{MIMONOMAvsOMA,Marcano}. Conventional NOMA allocates more power to the users with poor channel conditions to increase their throughput \cite{MIMONOMA}. Successive interference cancellation (SIC) is used at the receivers with better channel conditions to eliminate interference and decode their own signals. Similarly, IC benefits from operating in the very strong interference regime whenever possible as a design option. For that purpose, weak interference, which is determined based on the noise power, should be allowed and strong interference should be shifted to the very strong interference regime through power allocation.

Parallel Gaussian IC (PGIC) is a special case of multiple-input multiple-output (MIMO) system, where all the channel matrices are diagonal. In this paper, we consider a PGIC with two transmitter and receiver pairs and two subchannels, and derive a \emph{generalized iterative waterfilling algorithm} for power allocation that maximizes the ergodic achievable rate. This scheme enables NOMA by either allowing weak interference or intentionally forcing each transmitter to \emph{create very strong interference at the receiver and in each subchannel}. In particular, strong interference is shifted to the very strong interference regime by allocating the right power level at each transmitter and in each subchannel. Then very strong interference can be completely canceled at the receivers via SIC. We compare our approach to two other schemes when the interference is treated as noise using either optimal or sub-optimal power allocation. The results show that the proposed scheme significantly improves the sum-rate of PGIC channel.  

The rest of this paper is organized as follows. Section \ref{sec:problem_definition} presents the system model and describes the optimization problem for rate maximization. Section \ref{sec:waterfilling} presents the power allocation with generalized iterative waterfilling algorithm. Section \ref{sec:simResults} provides the simulation results on sum-rate gains. Section \ref{sec:conclusion} concludes the paper.

\section{Problem Definition}
\label{sec:problem_definition}

\begin{figure}[h]
	\centering
			\includegraphics[width=0.48\textwidth,trim={0.3cm 5cm 0.3cm 3.6cm},clip]{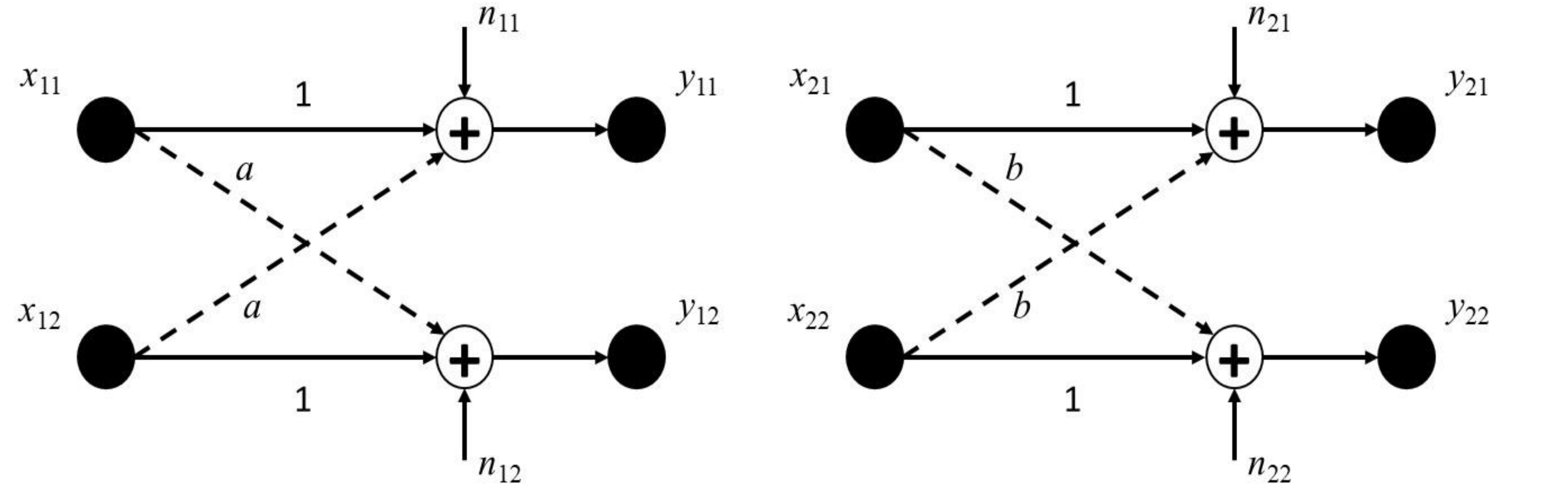}		
	\captionsetup{justification=centering}
	\caption{Parallel Gaussian interference channel with two users and two subchannels.} \label{fig:pCh}		
\end{figure}

We consider two transmitter and receiver pairs with two parallel Gaussian channels, i.e., these pairs communicate through two independent subchannels, as shown in Figure \ref{fig:pCh}. While there is no interference between the parallel subchannels, the interference exists due to multiple transmitters communicating on the same subchannel. Without loss of generality, we consider a symmetric channel with unit gain for the intended links. The channel gain is $a$ for the first subchannel and $b$ for the second subchannel for the interference links.  

The signal received at the first receiver in the first and second subchannels are given by

\begin{equation*}\label{eq:rxsignal1}
\begin{aligned}
y_{11}=\sqrt{P_{1}}x_{11}+\sqrt{Q_{1}a}x_{12}+\sqrt{N_{11}}n_{11}, \\
y_{21}=\sqrt{P_{2}}x_{21}+\sqrt{Q_{2}b}x_{22}+\sqrt{N_{21}}n_{21},
\end{aligned}
\end{equation*}
where $x_{11}$ and $x_{21}$ are the intended symbols, $P_1$ and $P_2$ are the power levels of the intended transmitters, $x_{12}$ and $x_{22}$ are the interference symbols, $Q_1$ and $Q_2$ are the power levels of the interference transmitters, and $N_{11}$ and $N_{21}$ are the noise variances at the first and second antenna elements of the first receiver, respectively.

The signal received at the second receiver in the first and second subchannels are given by 
\begin{equation*}\label{eq:rxsignal2}
\begin{aligned}
y_{12}=\sqrt{P_{1}a}x_{11}+\sqrt{Q_{1}}x_{12}+\sqrt{N_{12}}n_{12},\\
y_{22}=\sqrt{P_{2}b}x_{21}+\sqrt{Q_{2}}x_{22}+\sqrt{N_{22}}n_{22},
\end{aligned}
\end{equation*}
where $x_{11}$ and $x_{21}$ are the interference symbols, $P_1$ and $P_2$ are the power levels of the interference transmitters, $x_{12}$ and $x_{22}$ are the intended symbols, $Q_1$ and $Q_2$ are the power levels of the intended transmitters, and $N_{12}$ and $N_{22}$ are the noise variances at the first and second antenna elements of the second receiver, respectively.

We assume that the noise variance at each receiver is the same for both subchannels and equal to $N_{11} = N_{12} = N_{21} = N_{22} = \sigma^{2}$. Very strong interference occurs when the channel gains and allocated power levels satisfy the following conditions \cite{Carleial}:

\begin{eqnarray*}
&&a \geq \frac{P_1}{\sigma^{2}}+1 \text{ or } a \geq \frac{Q_1}{\sigma^{2}}+1 \text{ for subchannel 1,} \\
&&b \geq \frac{P_2}{\sigma^{2}}+1 \text{ or } b \geq \frac{Q_2}{\sigma^{2}}+1 \text{ for subchannel 2.}
\end{eqnarray*}

Strong interference occurs when $a \geq 1$ for subchannel $1$ and $b \geq 1$ for subchannel $2$, and weak interference occurs when $a < 1$ for subchannel $1$ and $b < 1$ for subchannel $2$. Interference regime is determined independent of the allocated transmit power for weak interference.

\begin{figure}[h]
\centering
\includegraphics[width=0.3\textwidth,trim={6cm 1.7cm 6cm 2.3cm},clip]{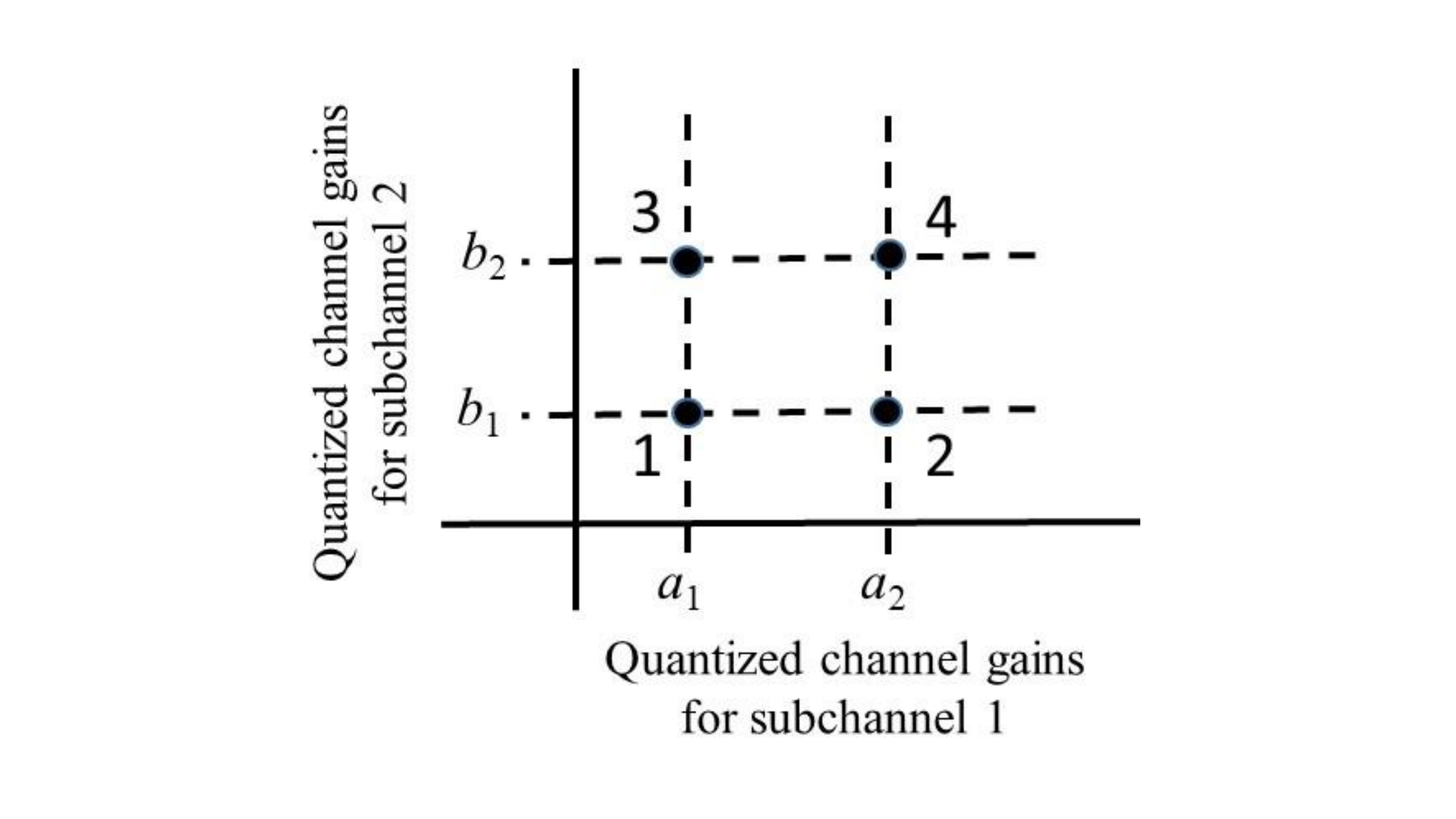}
\caption{Quantized channel gains that represent weak and strong/very strong interference regimes.}
\label{fig:ChGrid}
\end{figure}

Initially, the receivers estimate the channel gains corresponding to intended transmissions and interference. Suppose $a_1$ and $b_1$ are assigned as the channel gains whenever the estimated channel gain causes weak interference, i.e., $a_1 < 1$ and $b_1 < 1$, and $a_2$ and $b_2$ are assigned as the channel gains whenever the estimated channel gain causes either strong or very strong interference (depending on the allocated power), i.e., $a_2 \geq 1$ and $b_2 \geq 1$, as shown in Figure \ref{fig:ChGrid}. The numbers $1$, $2$, $3$, and $4$ on the grid correspond to transitions across \textit{weak-weak}, \textit{strong/very strong-weak}, \textit{weak-strong/very strong} and \textit{strong/very strong-strong/very strong} interference regimes, respectively.   

Our goal is to allocate the right power levels to the transmitters to create very strong interference at the receivers whenever the channel gain corresponds to $a_2$ or $b_2$. This approach will allow SIC at the receivers. 

We formulate an optimization problem given by (\ref{eq:Obj})-(\ref{eq:Power2}) to maximize the ergodic rate. In this formulation, the maximum PGIC rate over a number of channel realizations is provided in (\ref{fig:cost}), where $P_{ij}$ corresponds to $i$th subchannel and $j$th interference regime for the first transmitter, and $Q_{ij}$ corresponds to $i$th subchannel and $j$th interference regime for the second transmitter. Interference is treated as noise whenever it is weak (which is optimal). Interference is cancelled whenever it is very strong. This way, strong interference regime is forced to shift to the very strong interference regime with allocated power. Constraints (\ref{eq:Pconst}) and (\ref{eq:Qconst}) set the maximum values on sum-power for all the instantaneous powers used by a transmitter on both of the subchannels over time. The power levels need to be greater than or equal to $0$ at any given time. Constraints (\ref{eq:Power1}) and (\ref{eq:Power2}) need to be satisfied to create very strong interference at the receivers.  

\begin{figure*}[!t]
	\normalsize
	\setcounter{MYtempeqncnt}{\value{equation}}
	\setcounter{equation}{\value{MYtempeqncnt}}
	\begin{IEEEeqnarray}{rCl} \label{fig:cost} 
	& C = &\max 
	\Bigg( P(a=a_1, b=b_1)C_{11} 
 + P(a=a_2, b=b_1)C_{21}
 + P(a=a_1, b=b_2)C_{12} 
 + P(a=a_2, b=b_2)C_{22} \Bigg),  \label{eq:Obj} \\
& \textnormal{where} \nonumber \\
&& C_{11} = \log_{2} \Bigg( 1+\frac{P_{11}}{\sigma^2+Q_{11}a_1} \Bigg) + \log_{2} \Bigg( 1+\frac{P_{21}}{\sigma^2+Q_{21}b_1}\Bigg) + \log_{2} \Bigg( 1+\frac{Q_{11}}{\sigma^2+P_{11}a_1} \Bigg) + \log_{2} \Bigg( 1+\frac{Q_{21}}{\sigma^2+P_{21}b_1}\Bigg), \nonumber \\
&& C_{21} = \log_{2} \Bigg( 1+\frac{P_{12}}{\sigma^2} \Bigg) + \log_{2} \Bigg( 1+\frac{P_{22}}{\sigma^2+Q_{22}b_1} \Bigg) + 
	\log_{2} \Bigg( 1+\frac{Q_{12}}{\sigma^2} \Bigg) + \log_{2} \Bigg( 1+\frac{Q_{22}}{\sigma^2+P_{22}b_1} \Bigg),\nonumber \\
&& C_{12} = \log_{2} \Bigg( 1+\frac{P_{13}}{\sigma^2+Q_{13}a_1} \Bigg) + \log_{2} \Bigg( 1+\frac{P_{23}}{\sigma^2} \Bigg) + 
	\log_{2} \Bigg( 1+\frac{Q_{13}}{\sigma^2+P_{13}a_1} \Bigg) + \log_{2} \Bigg( 1+\frac{Q_{23}}{\sigma^2} \Bigg), \nonumber \\	
&& C_{22} = \log_{2} \Bigg( 1+\frac{P_{14}}{\sigma^2} \Bigg) + \log_{2} \Bigg( 1+\frac{P_{24}}{\sigma^2} \Bigg) + \log_{2} \Bigg( 1+\frac{Q_{14}}{\sigma^2} \Bigg) + \log_{2} \Bigg( 1+\frac{Q_{24}}{\sigma^2} \Bigg) \nonumber \\
& \textnormal{subject to} \nonumber \\
&& P_{11}+P_{12}+P_{13}+P_{14}+P_{21}+P_{22}+P_{23}+P_{24} \leq P, \hspace{4em}  P_{ij} \geq 0 \textnormal{ for } i=1,2 \textnormal{ and } j=1,2,3,4, \label{eq:Pconst} \\
&& Q_{11}+Q_{12}+Q_{13}+Q_{14}+Q_{21}+Q_{22}+Q_{23}+Q_{24} \leq Q, \hspace{3em}  Q_{ij} \geq 0 \textnormal{ for } i=1,2 \textnormal{ and } j=1,2,3,4, \label{eq:Qconst} \\
&& P_{12} \leq a_2 \sigma^2 - \sigma^2, \hspace{3em} P_{14} \leq a_2 \sigma^2 - \sigma^2, \hspace{3em} Q_{13} \leq a_2 \sigma^2 - \sigma^2, \hspace{3em} Q_{14} \leq a_2 \sigma^2 - \sigma^2, \label{eq:Power1}\\ 
&& P_{22} \leq b_2 \sigma^2 - \sigma^2,  \hspace{3em} P_{24} \leq b_2 \sigma^2 - \sigma^2, \hspace{3em} \: Q_{23} \leq b_2 \sigma^2 - \sigma^2, \hspace{3em} \: Q_{24} \leq b_2 \sigma^2 - \sigma^2. \label{eq:Power2}  
\end{IEEEeqnarray}
\setcounter{equation}{\value{MYtempeqncnt}}
\hrulefill
\vspace*{4pt} 
\end{figure*}

We solve the underlying optimization problem and determine the maximum achievable rate. For that purpose, we derive the optimal power allocation that maximizes the sum-rate for given channel gains offline and then assign them to each antenna during real-time operation based on the estimated interference regimes.

\section{Power Allocation with Generalized Iterative Waterfilling Algorithm}
\label{sec:waterfilling}

The cost function defined in (\ref{fig:cost}) is the sum of concave and convex terms with convex constraint sets (\ref{eq:Pconst})-(\ref{eq:Power2}). The solution to the maximization problem in (\ref{fig:cost}) should satisfy the Karush-Kuhn-Tucker (KKT) conditions, which would ensure a local minimum. The Lagrangian is given in (\ref{eq:lagrangian}). 

\setcounter{equation}{5}

\begin{eqnarray} \label{eq:lagrangian}
\begin{aligned}  
&\Lagr (P, Q; \lambda, \mu, \bm{\psi}, \bm{\varepsilon}, \bm{\omega}, \bm{\zeta}) = \\  & - \Bigg(P(a=a_1, b=b_1) C_{11} + P(a=a_2, b=b_1) C_{21} \\ & + P(a=a_1, b=b_2) C_{12} + P(a=a_2, b=b_2) C_{22} \Bigg) \\
& + \lambda \Bigg(\sum_{i=1}^2 \sum_{j=1}^4 P_{ij} - P\Bigg) + \mu\Bigg(\sum_{i=1}^2 \sum_{j=1}^4 Q_{ij} - Q\Bigg) \\& - \sum_{i=1}^2 \sum_{j=1}^4 \psi_{ij}P_{ij}-\sum_{i=1}^2 \sum_{j=1}^4 \varepsilon_{ij}Q_{ij} \\ & + 
 \omega_1(P_{12}-a_2\sigma^2+\sigma^2)+\omega_2(P_{22}-b_2\sigma^2+\sigma^2) \\ & +
\omega_3(P_{14}-a_2\sigma^2+\sigma^2)+\omega_4(P_{24}-b_2\sigma^2+\sigma^2)\\ & + 
 \zeta_1(Q_{13}-a_2\sigma^2+\sigma^2)+\zeta_2(Q_{23}-b_2\sigma^2+\sigma^2)\\ & +
\zeta_3(Q_{14}-a_2\sigma^2+\sigma^2)+\zeta_4(Q_{24}-b_2\sigma^2+\sigma^2).
\end{aligned}
\end{eqnarray}

The KKT conditions are given by 
\begin{eqnarray}
\lambda \geq 0, \mu \geq 0, \psi_{ij} \geq 0, \varepsilon_{ij} \geq 0, \omega_{j} \geq 0, \zeta_{j} \geq 0, \nonumber 
\end{eqnarray} 
 
\begin{eqnarray} 
\lambda \Bigg(\sum_{i=1}^2 \sum_{j=1}^4 P_{ij} - P\Bigg)=0,  \:\: \mu\Bigg(\sum_{i=1}^2 \sum_{j=1}^4 Q_{ij} - Q\Bigg)=0, \nonumber \end{eqnarray}
\begin{eqnarray} 
\psi_{ij}P_{ij}=0, \: \varepsilon_{ij}Q_{ij}=0, \text{for } i=1,2 \text{ and } j=1,2,3,4,\nonumber 
\end{eqnarray}
\begin{eqnarray} 
\omega_1(P_{12}-a_2\sigma^2+\sigma^2)=0, \: & \omega_2(P_{22}-b_2\sigma^2+\sigma^2)=0, \nonumber \\
\omega_3(P_{14}-a_2\sigma^2+\sigma^2)=0, \: & \omega_4(P_{24}-b_2\sigma^2+\sigma^2)=0, \nonumber \\
\end{eqnarray}

\begin{eqnarray} 
\zeta_1(Q_{13}-a_2\sigma^2+\sigma^2)=0, \: &  \zeta_2(Q_{23}-b_2\sigma^2+\sigma^2)=0, \nonumber \\
\zeta_3(Q_{14}-a_2\sigma^2+\sigma^2)=0, \: & \zeta_4(Q_{24}-b_2\sigma^2+\sigma^2)=0. \nonumber 
\end{eqnarray}

The optimum power allocation policy should simultaneously satisfy all these KKT conditions. Moreover, the optimum power value of each transmitter depends on the power levels  assigned to both transmitters and subchannels in different interference regimes. The derivative terms that include quadratic power and noise terms should be solved jointly. It is hard to analytically solve for the optimum power allocations from the KKT conditions. We treat this problem as a \emph{generalized waterfilling algorithm}, where the water levels, $1/\lambda$ and $1/\mu$, are the sum of quadratic power and noise terms. 

We follow an iterative algorithm to solve this problem. First, all power levels $P_{ij}$ and $Q_{ij}$ for $i=1,2$ and $j=1,2,3,4$ are initialized. Then we iteratively update the power levels of the first transmitter ($P_{ij}$) by assuming the power levels of the second transmitter ($Q_{ij}$) are fixed, and then update $Q_{ij}$s by assuming $P_{ij}$s are fixed. To find the optimal $P_{ij}$s, generalized waterfilling algorithm is derived to jointly optimize the power levels by initially setting the water level as 
\begin{eqnarray} 
\begin{aligned}
\frac{1}{\lambda} = \min\Bigl(f\left(P_{11}\right),f\left(P_{12},\omega_1\right),
f\left(P_{13}\right),f\left(P_{14},\omega_2\right),\\
f\left(P_{21}\right),f\left(P_{22},\omega_3\right),f\left(P_{23}\right),f\left(P_{24},\omega_4\right)\Bigr),\nonumber
\end{aligned}
\end{eqnarray}
where 
\begin{eqnarray}
\begin{aligned}
&f(P_{11}) = \\
& \frac{1}{\sigma^2+Q_{11}a_1+P_{11}} + \frac{a_1}{\sigma^2+P_{11}a_1+Q_{11}}-\frac{a_1}{\sigma^2+P_{11}a_1},\\
&f(P_{12},\omega_1) = \frac{1}{\sigma^2+P_{12}}-\omega_1, \nonumber \\
\end{aligned}
\end{eqnarray}

\begin{eqnarray}
\begin{aligned}
&f(P_{13}) = \\
& \frac{1}{\sigma^2+Q_{13}a_1+P_{13}} + \frac{a_1}{\sigma^2+P_{13}a_1+Q_{13}}-\frac{a_1}{\sigma^2+P_{13}a_1},\\
&f(P_{14},\omega_2) = \frac{1}{\sigma^2+P_{14}}-\omega_2,\\
&f(P_{21}) = \\ 
&\frac{1}{\sigma^2+Q_{21}b_1+P_{21}} + \frac{b_1}{\sigma^2+P_{21}b_1+Q_{21}}-\frac{b_1}{\sigma^2+P_{21}b_1},\\
&f(P_{22},\omega_3) = \frac{1}{\sigma^2+Q_{22}b_1+P_{22}} + \\
&\frac{b_1}{\sigma^2+P_{22}b_1+Q_{22}}-\frac{b_1}{\sigma^2+P_{22}b_1}-\omega_3,\\
&f(P_{23}) = \frac{1}{\sigma^2+P_{23}},  \:\:\: f(P_{24},\omega_4) = \frac{1}{\sigma^2+P_{24}}-\omega_4.\nonumber 
\end{aligned}
\end{eqnarray}
The roots of the quadratic equations provided above, which are the power values of the first transmitter, are computed. Initially, $\omega_j$s, where $j=1,2,3,4$, are set to $0$ and their values are optimized in the algorithm to maximize the sum-rate. After determining the initial water level, this iterative process of increasing the water level and calculating the allocated power for each interference regime continues until all the available power, i.e., $P$, is consumed. The pseudo code to solve this optimization problem is presented in Algorithm \ref{alg:waterfilling1}.

\begin{algorithm}[H]
\caption{Generalized iterative waterfilling algorithm to compute $P_{ij}s$. }\label{alg:waterfilling1}
\begin{algorithmic}[1]
\Procedure{Allocate power levels based on the interference regime}{}
\State Set $\omega_j=0$ and initialize $1/\lambda$ as $\min(f(P_{ij}))$;
\State Set the tolerance as $\tau$;
\While{$(P-\sum_{i=1}^2\sum_{j=1}^4P_{ij}) > \tau$}
\State Calculate the corresponding power levels for $1/\lambda$ while iterating over $\omega_j$s; 
\State Select $P_{ij}=\max(P_{ij},0)$ that maximizes rate for given $\omega_j$s;
\State Calculate the updated water level, $(1/\lambda)_{new}$;
\If{$(1/\lambda)_{new}\neq(1/\lambda)$}
\State Set $1/\lambda = (1/\lambda)_{new}$
\Else
\State Increase the water level, $1/\lambda$.
\EndIf
\EndWhile
\EndProcedure
\end{algorithmic}
\end{algorithm}

Next, using the optimized power levels for the first transmitter, the power levels for the second transmitter are optimized.

Generalized waterfilling algorithm is used to find the optimal $Q_{ij}$s by setting the water level as
\begin{eqnarray} 
\begin{aligned}
\frac{1}{\mu} = \min\Bigl(f\left(Q_{11}\right),f\left(Q_{12}\right),
f\left(Q_{13},\zeta_1\right),f\left(Q_{14},\zeta_2\right),\\
f\left(Q_{21}\right),f\left(Q_{22}\right),f\left(Q_{23},\zeta_3\right),f\left(Q_{24},\zeta_4\right)\Bigr), \nonumber
\end{aligned}
\end{eqnarray}
where 
\begin{eqnarray*}
\begin{aligned}
&f(Q_{11}) = \\
&\frac{a_1}{\sigma^2+Q_{11}a_1+P_{11}} - \frac{a_1}{\sigma^2+Q_{11}a_1}+\frac{1}{\sigma^2+P_{11}a_1+Q_{11}},\\
\end{aligned}
\end{eqnarray*}

\begin{eqnarray*}
\begin{aligned}
&f(Q_{12}) = \frac{1}{\sigma^2+Q_{12}},\\
&f(Q_{13},\zeta_1) = \frac{a_1}{\sigma^2+Q_{13}a_1+P_{13}} - \frac{a_1}{\sigma^2+Q_{13}a_1}+\\
&\frac{1}{\sigma^2+P_{13}a_1+Q_{13}}-\zeta_1,\\
&f(Q_{14},\zeta_2) = \frac{1}{\sigma^2+Q_{14}}-\zeta_2,\\
&f(Q_{21}) = \\
& \frac{b_1}{\sigma^2+Q_{21}b_1+P_{21}} - \frac{b_1}{\sigma^2+Q_{21}b_1}+\frac{1}{\sigma^2+P_{21}b_1+Q_{21}},\\
&f(Q_{22}) = \\
& \frac{b_1}{\sigma^2+Q_{22}b_1+P_{22}} - \frac{b_1}{\sigma^2+Q_{22}b_1}+\frac{1}{\sigma^2+P_{22}b_1+Q_{22}},\\
&f(Q_{23},\zeta_3) = \frac{1}{\sigma^2+Q_{23}}-\zeta_3, \: f(Q_{24},\zeta_4) = \frac{1}{\sigma^2+Q_{24}}-\zeta_4.
\end{aligned}
\end{eqnarray*}

The roots of the quadratic equations provided above, which are the power values of the second transmitter, are computed. The pseudo code to solve this optimization problem is the same as Algorithm \ref{alg:waterfilling1} with $\omega_j$, $\lambda$, $P_{ij}$ and $P$ replaced by $\zeta_j$, $\mu$, $Q_{ij}$ and $Q$, respectively. The power values for both of the transmitters are calculated iteratively until the results converge according to the tolerance level, $\tau$, as defined in Algorithm \ref{alg:waterfilling1}.

\section{Performance Evaluation}
\label{sec:simResults}

We simulated the generalized waterfilling algorithm using Matlab. The following default parameters are used. $a_1$ and $b_1$ are set to $0.1$ that represent weak interference and $a_2$ and $b_2$ are set to $10$ that represent strong/very strong interference. $P$ and  $Q$ are set to $50$ (where noise power is normalized to $1$). $\tau$ is set to $10^{-5}$. The channel realization is repeated to generate an equal number of \textit{weak-weak}, \textit{weak-strong/very strong}, \textit{strong/very strong-weak} and \textit{strong/very strong-strong/very strong} channel realizations in two subchannels, i.e., $P(a=a_1, b=b_1) = P(a=a_2, b=b_1) = P(a=a_1, b=b_2) =  P(a=a_2, b=b_2) = 1/4$. 

We compare the sum-rate under three schemes:
\underline{Scheme 1}: NOMA with optimal power allocation is the proposed scheme. The sum-rate is calculated as $10.4$ b/s/Hz, when very strong interference is created at the receivers whenever the channel gain is higher than $1$. 

\underline{Scheme 2}: Interference is treated as additional noise in the rate formulation. We solve the rate maximization in an iterative way as described above by first optimizing the power levels for the first transmitter and then for the second transmitter. The iterative waterfilling algorithm allocates power levels close to $0$ to the first transmitter whenever the channel gains are higher than $1$. As a result, only one transmitter becomes active on the channel when the interference regime is strong/very strong. The second transmitter achieves higher throughput when the allocated power is minimal at the first transmitter. The sum-rate is calculated as $8$ b/s/Hz.

\underline{Scheme 3}: Total power is always divided equally for every channel realization independent of the interference regime and then the interference is treated as noise. The sum-rate is calculated as $4.8$ b/s/Hz for this scheme.   

\begin{figure}
	\centering
	\subfloat[]{\includegraphics[width= 0.48\columnwidth,trim={0.6cm 0cm 0.8cm 0.5cm},clip]{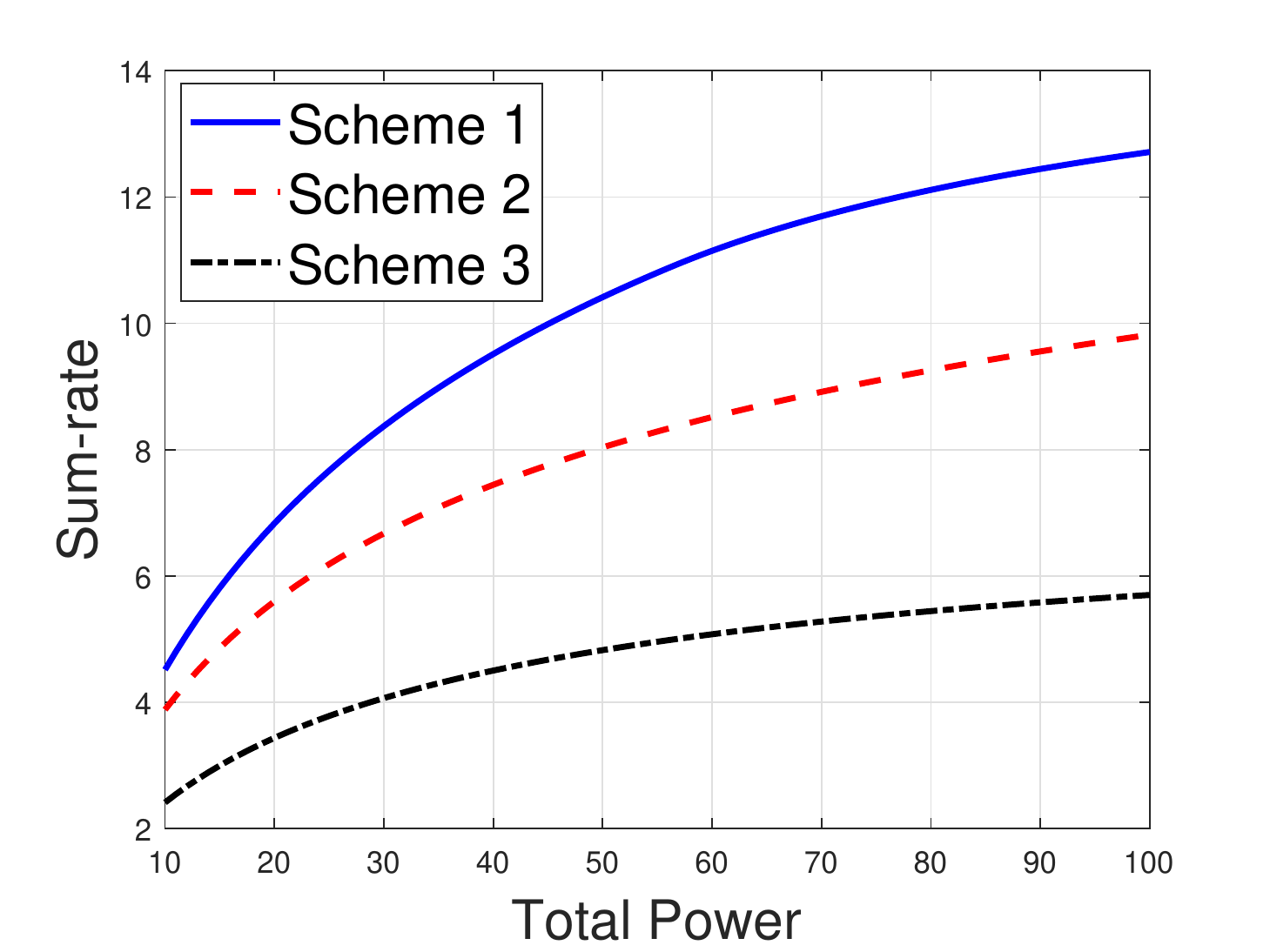}}
\subfloat[]{\includegraphics[width= 0.48\columnwidth,trim={0.6cm 0cm 0.8cm 0.5cm},clip]{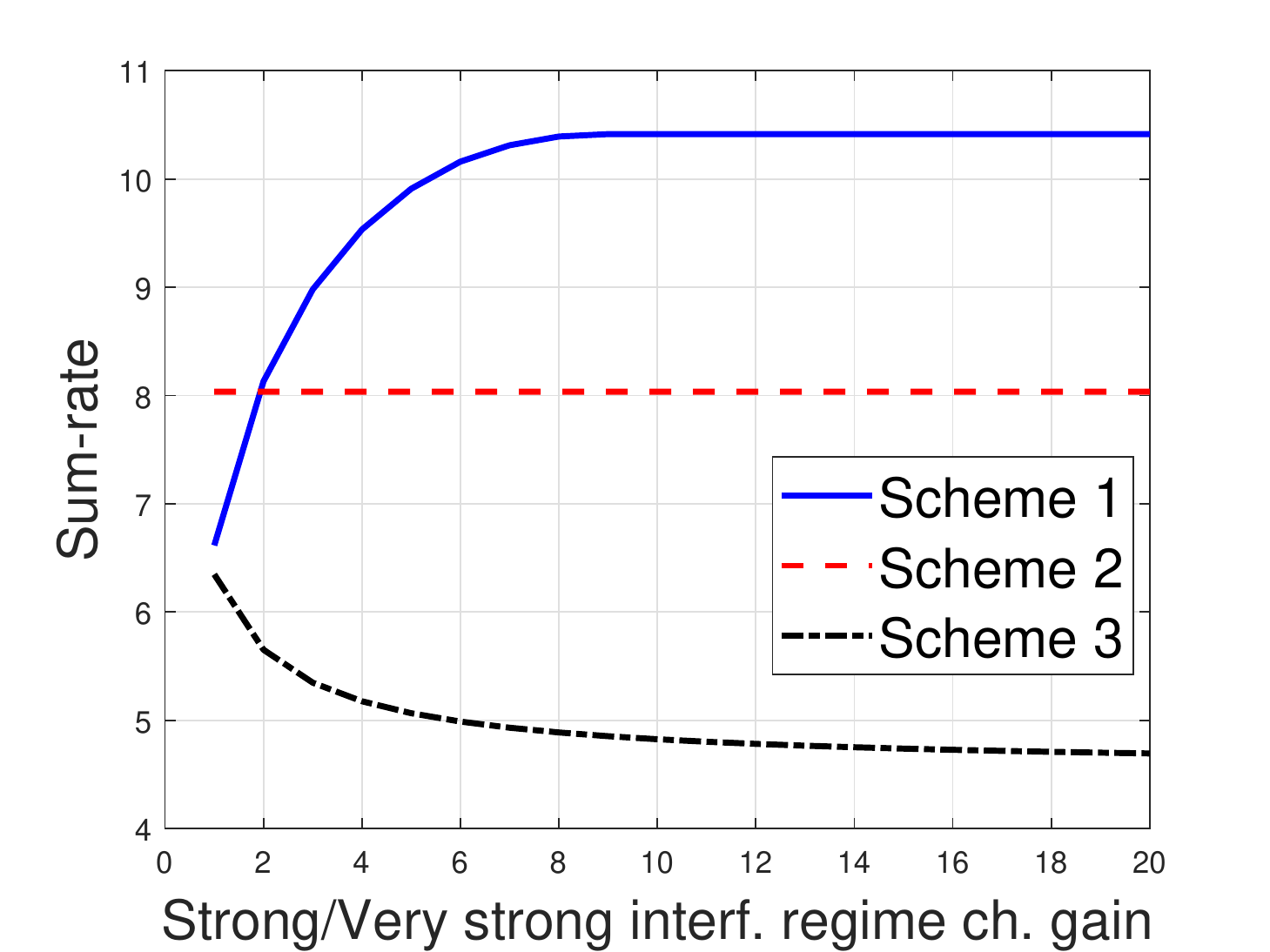}}
	\captionsetup{justification=centering}
	\caption{Sum-rate for (a) power levels varying between $10$ and $100$, (b) strong/very strong interference quantized channel gains, $a_2$ and $b_2$, varying between $1$ and $20$.} \label{fig:compSumRateChGain}
\end{figure}

Next, we simulate the sum-rate by varying total power $P$ and  $Q$  between $10$ and $100$. The rest of the simulation parameters are kept the same as before. Figure \ref{fig:compSumRateChGain} (a) shows the simulation results. Scheme 1 improves the sum-rate of the PGICs significantly (up to $123.1\%$ compared with Scheme 3 and $29.5\%$ compared with Scheme 2).

We also calculate the sum-rate with strong/very strong interference quantized channel gain values, $a_2$ and $b_2$, varying from $1$ to $20$. The results shown in Figure \ref{fig:compSumRateChGain} (b) provide insight on determining the quantized channel gain values for strong/very strong interference regime. Schemes 1-3 are executed as above with $P = Q = 50$ and by varying $a_2$ and $b_2$. When $a_2 = b_2 = 1$, Scheme 1 performs worse compared to Scheme 2, since no power is allocated on the subchannels whenever the channel gain is equal to $1$ to satisfy the very strong interference constraints. The sum-rate continues to increase with increasing quantized channel gains and tapers off when $a_2=b_2=8$. For Scheme 2, the sum-rate is independent of  $a_2$ and $b_2$, since KKT equations yield zero power allocated for the first transmitter when there is no interference cancellation. This zero power eliminates the corresponding signal-to-interference-and-noise-ratio (SINR) term in the rate expression when it is in the numerator and eliminates the $a_2$ and $b_2$ terms when it is in the denominator. The rate decreases with increasing interference levels for Scheme 3, since the power is always divided equally. Scheme 1 improves the sum-rate of the PGICs significantly (up to $121.9\%$ compared with Scheme 3 and $29.6\%$ compared with Scheme 2). 

\begin{figure}
	\centering
	\subfloat[]{\includegraphics[width= 0.48\columnwidth,trim={0.6cm 0cm 0.8cm 0.5cm},clip]{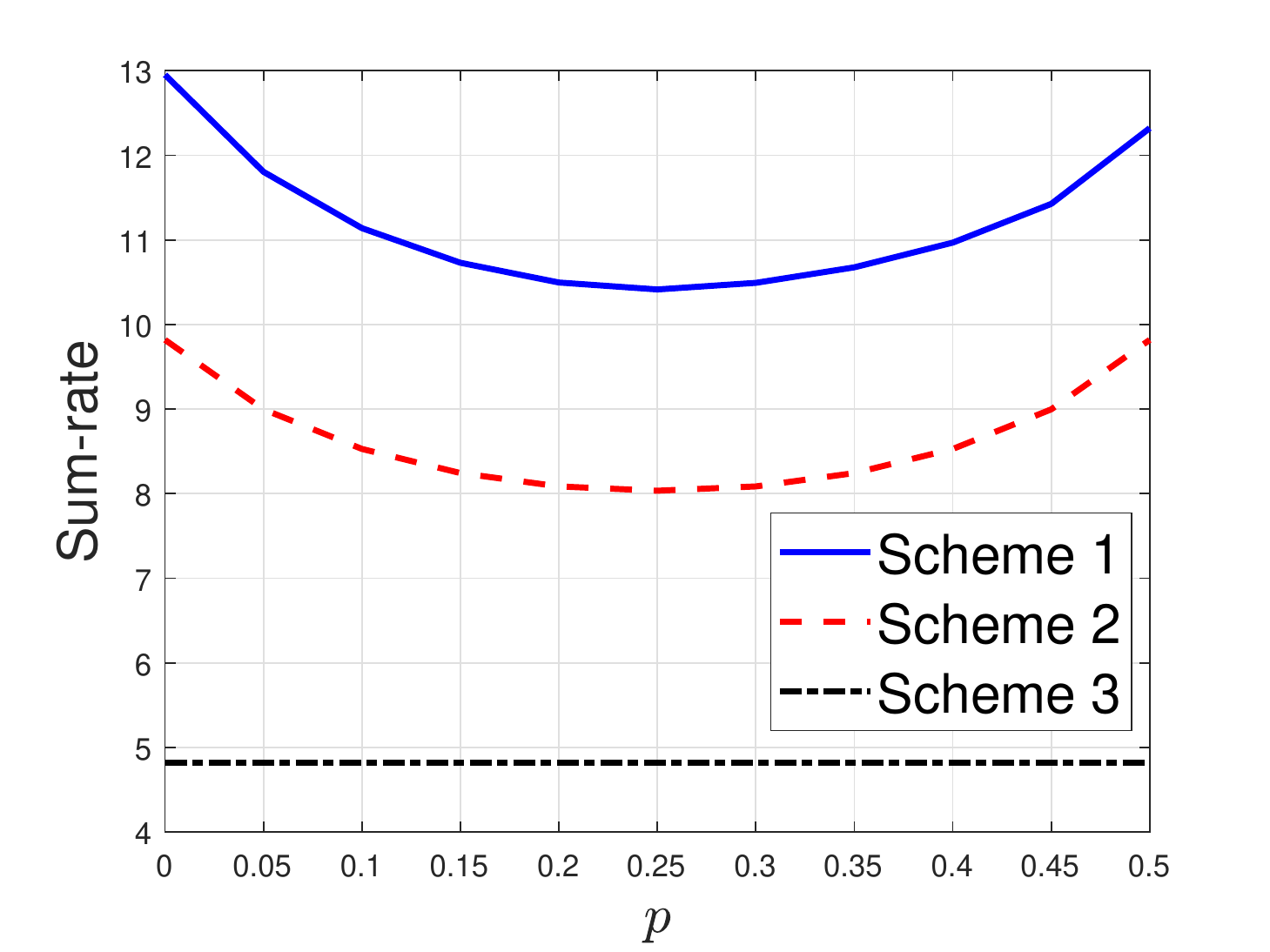}}
\subfloat[]{\includegraphics[width= 0.48\columnwidth,trim={0.6cm 0cm 0.8cm 0.5cm},clip]{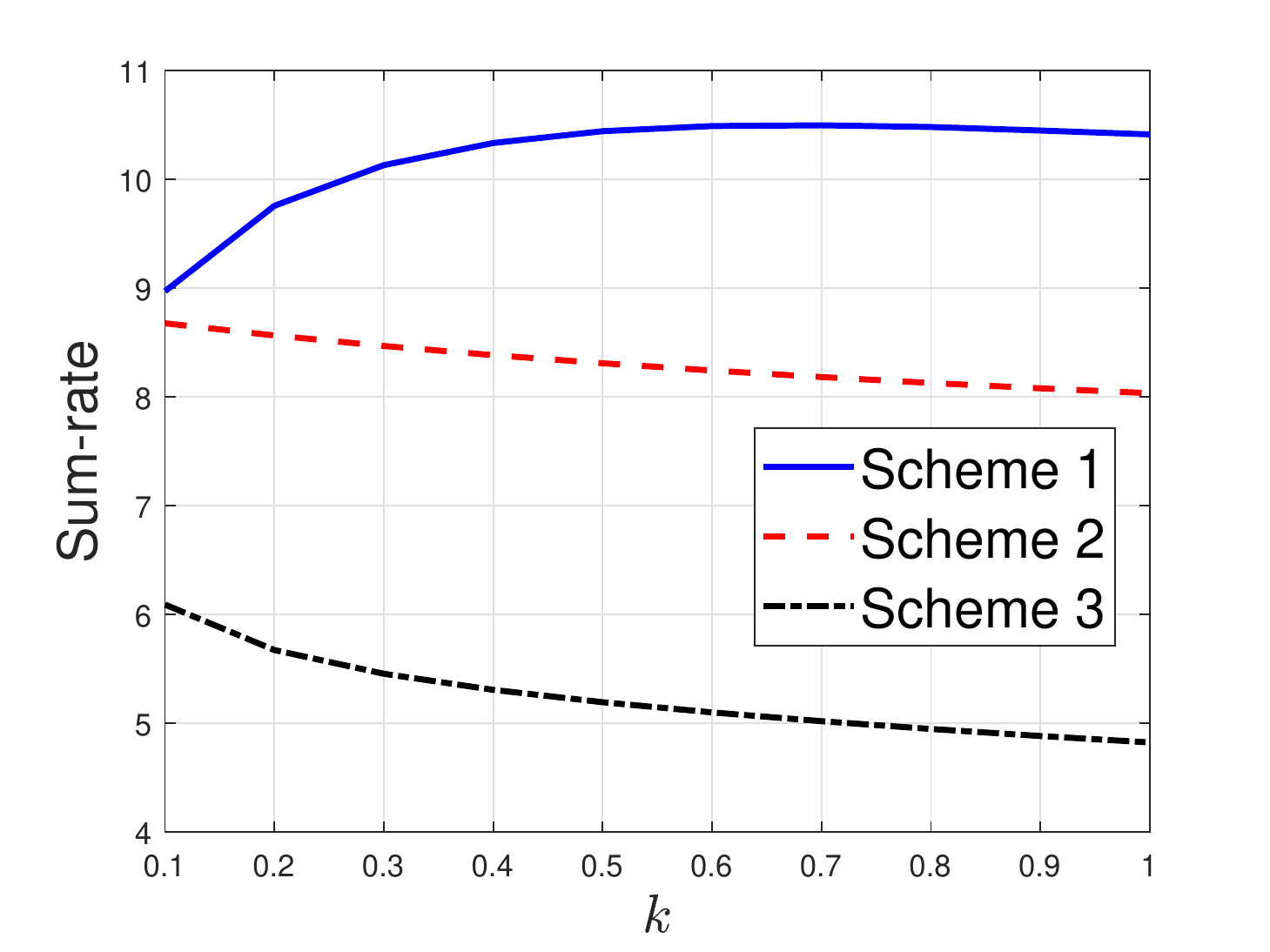}}
	\captionsetup{justification=centering}
	\caption{Sum-rate for (a) different probability distributions for interference regimes, (b) different probability distributions for asymmetrical channel gains.} \label{fig:compProbAssym}
\end{figure}

Next, we simulate the effect of probability distributions assigned to interference regimes on the sum-rate by varying the probability $p$ from $0$ to $0.5$ and assigning it as  $P(a=a_1, b=b_1)=P(a=a_2, b=b_2)=p$, $P(a=a_1, b=b_2)=P(a=a_2, b=b_1)=0.5-p$. Figure \ref{fig:compProbAssym} (a) shows the sum-rate as a function of $p$. The probability distributions for Scheme 3 remain the same with varying $p$, since $P$s and $Q$s are always allocated equally for each transmitter and subchannel. Scheme 1 outperforms Scheme 2 and Scheme 3 for all $p$ values.  Scheme 1 improves the sum-rate of the PGICs significantly (up to $168.4\%$ compared with Scheme 3 and $31.9\%$ compared with Scheme 2).  

Finally, we evaluate the sum-rate under asymmetrical channel gains in subchannels. We set $a_1=0.1$, $a_2=10$, $b_1=k \times a_1$, $b_2=k \times a_2$, where $k$ varies from $0.1$ to $1$, $P = 50$, $Q = 50$, and $p = 1/4$. When $k=1$, the results correspond to our default setting with $a_1=b_1=1$ and $a_2=b_2=10$. The results in Figure \ref{fig:compProbAssym} (b) show that Scheme 1 improves the sum-rate of the PGICs significantly (up to $115.8\%$ compared with Scheme 3 and $29.6\%$ compared with Scheme 2).  

\section{Conclusion}
\label{sec:conclusion}

We developed the interference regime enforcing scheme with the iterative waterfilling that allocates power levels for PGICs to enable NOMA and maximize the achievable rate. This scheme creates very strong interference at the receivers by power allocation whenever the channel gain is larger than the ratio of the noise variances at the receivers operating in the same subchannel. Then SIC is applied at the receivers. Results showed that this scheme provides major improvement to the sum-rate compared with two other benchmark schemes when interference is treated as noise only with optimal or sub-optimal power allocation. 

\bibliographystyle{ieeetr}

\end{document}